\documentclass[conference]{IEEEtran}
\IEEEoverridecommandlockouts
\usepackage{cite}
\usepackage{url}
\usepackage{amsmath,amssymb,amsfonts}
\usepackage{algorithmic}
\usepackage{graphicx}
\usepackage{textcomp}
\usepackage{xcolor}
\usepackage{hyperref}
\usepackage{booktabs}
\usepackage{graphicx}
\usepackage{longtable}
\usepackage{dcolumn}
\usepackage{float}
\usepackage[caption=false]{subfig}

\def\BibTeX{{\rm B\kern-.05em{\sc i\kern-.025em b}\kern-.08em
    T\kern-.1667em\lower.7ex\hbox{E}\kern-.125emX}}
\begin{document}

\title{A systematic literature review on insider threats}

\author{\IEEEauthorblockN{Angad Pal Singh}
\IEEEauthorblockA{\textit{School of Engineering} \\
\textit{University of Guelph}\\
Guelph, Canada \\
angadpal@uoguelph.ca}

\and
\IEEEauthorblockN{Ankit Sharma}
\IEEEauthorblockA{\textit{School of Engineering} \\
\textit{University of Guelph}\\
Guelph, Canada \\
asharm50@uoguelph.ca}

}

\maketitle

\begin{abstract}
Insider threats is the most concerned cybersecurity problem which is poorly addressed by widely used security solutions. Despite the fact that there have been several scientific publications in this area, but from our innovative study classification and structural taxonomy proposals, we argue to provide the more information about insider threats and defense measures used to counter them. While adopting the current grounded theory method for a thorough literature evaluation, our categorization's goal is to organize knowledge in insider threat research. Along with an analysis of major recent studies on detecting insider threats, the major goal of the study  is to develop a classification of current types of insiders, levels of access, motivations behind it, insider profiling, security properties, and methods they use to attack. This includes use of machine learning algorithm, behavior analysis, methods of detection and evaluation. Moreover, actual incidents related to insider attacks have also been analyzed \\ 


\end{abstract}

\section{Introduction\textbf{}}
\label{Introduction}

Today’s digital economy depends on strong cyber security. As opposed to external threats, internal threats to security are the biggest. Important mitigating approaches include the detection and forecasting of insider threats \cite{a3}. In the digital era in which we live has both its benefits and drawbacks, just like anything else. The biggest drawback is the security risk. Data breaches are becoming more frequent and important as our sensitive information migrates more toward the digital domain. A case in point would be businesses may be exposed to security hazards from both internal and external sources. For instance, internal attacks are more dangerous than external ones, whether they originate from employees, vendors, or other companies due to having a direct relationship and access to a company’s computer system  \cite{o1}. Because they are familiar with how the organization functions on a daily basis, these insiders have access to all the privileges and permissions necessary to start an attack that is not available to outsiders.

Insiders are able to pass off their attacks as normal operations which makes it difficult to distinguish them  \cite{o2,a2}.
A countermeasure of creating automated threat detection systems that don’t raise too many false alarms is the issue at hand. Employees might not be able to access the system during a time-sensitive situation and carry out their tasks successfully in emergency-like situations. In such situations, companies can become paralyzed by the loss of system availability, which can increase expenses, reduce income and pose real-life threats. One of the most recent case examples to share would be the one where 27,000 of client records were exposed by Barclays bank claimed by a whistleblower employee \cite{o3}. Not only did the leak result in loss of trust of the customers, but also resulted in £7.7 million in penalties and ordered to pay impacted parties up to £59 million in compensation. This is one of many incidents that occurred due to insider threats and it is of paramount importance to explore methods to prevent them from happening in today’s world \cite{a4,a5}.

1.1.	Prior research

To our knowledge, there appears to be very little systematic literature review(SLR), particularly with relation to insider threats in the modern world. One of the latest investigative papers in the field of cyber security and insider threats was by Al-Mhiqani, Mohammed and Ahmad, and various another writer in 2020 \cite{o4}. In this study, the authors highlight classification of contemporary insider types, access rights, levels, motives, insider profiles, security properties of validity and methods used by attackers. He also evaluated recent notable research on insider threat detection, including behavior modeling,and used implementing of machine learning algorithms
Several real cases of insider threats were analyzed to provide statistical information on insiders. Additionally, this survey highlights challenges other researchers face and makes recommendations to reduce barriers.

This section discusses insider threat investigations and review articles. Only a handful of documents were found, i.e., Walker-Roberts\cite{Ar1} published a review in 2018. He published a review paper on detecting insider threats; however, the scope of the paper was restricted to medical science. Another review by Ullah \cite{ar2} was published in 2018 identified and critically analyzed data intrusion attack vectors and countermeasures to report on the state of the art and identify vulnerabilities for future research. In 2011, famous Researchers Hunker and others summarized the main properties related to insiders and threats they cause to organization.\cite{o2}. The primary contributions of this review and how it differs from earlier studies can be summed up as follows:. First, to our knowledge, this is the first paper in which we have reviewed numerous papers related to insider threat and summarized a systematic report at the end of the study. Secondly, many incidents related to insiders has also been summarized and discussed. Moreover a summary and future work recommendation is also provided at the end of the paper. \cite{a1}.

1.2.	Research goals

   The main goal of this paper is to systematically review the available research papers and to analyze the Insider threat problem. This study addresses the following research questions that has been mentioned in Table \ref{tab:Table 1}.

1.3.	Contributions and layout

This SLR is addition to existing study and discuss the following contributions for those with an interest in the field of Cyber Security and threats to further their work:
\begin{table}[h]
\centering
\caption{Research Questions}
\label{tab:Table 1}
\scalebox{0.8}{
\begin{tabular}{@{}ll@{}}
\toprule
Research Questions (RQ) & Discussion   \\ \midrule
\begin{tabular}[c]{@{}l@{}}RQ1. What is the magnitude and impact \\ of insider threats to businesses globally?\end{tabular}                                                                    & \begin{tabular}[c]
{@{}l@{}}Insider threats has been increased since the \\introduction of internet, recent survey \\shows that insider  attack increased from 30 \% to \\ 51\% in couple of years.
\\
\end{tabular}             \\
\begin{tabular}[c]{@{}l@{}}RQ2. The common techniques used to\\  carry out insider threats and override \\ controls that are set to detect insider threats \\ in the first place.\end{tabular} & \begin{tabular}[c]{@{}l@{}}The most common insider threats \\ are caused by ex-employee or humans,\\  so to mitigate this companies have \\ adopted policy of changing or revising \\ their system after few years.\end{tabular} \\
\begin{tabular}[c]{@{}l@{}}RQ3. What methods and techniques are\\  available to overcome and insider threats \\ in the world of cyber security?\end{tabular}                                   & \begin{tabular}[c]{@{}l@{}}Educating our users and strong \\ authentications are the main key methods \\ to overcome this problem.\end{tabular}                                                                                 \\ \bottomrule
\end{tabular}}
\end{table}

•	We found 50 major studies on insider attacks and cyber security up to 2020. If other researchers desire to continue their work in this particular field, they are welcome to use our list of studies.

•	After that, we chose 37 primary studies that meet to the standards we established for quality evaluation. These studies can be used as helpful comparison points for other related research.

•	Following that, we made the decision to perform a thorough analysis of the data found in the subset of 37 studies. We then provided the data to illustrate the research idea related to the Threat Detection and it's implementation in cyber world.

•	We presented a conclusion at the end after analysis of all the studies.

•	We issue a statement and create guidelines for further support
work in this area.
\\
This essay is organised as follows: The techniques used to carefully choose the primary studies for analysis are described in the Section \ref{Research Methodology}
The conclusions of each of the chosen primary research are presented in Section \ref{Findings}. The findings in relation to the earlier posed study topics are discussed in Section \ref{Discussion}. Section \ref{Future Research Directions for Insider Threats} concludes the and makes some recommendations for more future work.
\section{Research Methodology}
\label{Research Methodology}
Our study’s objective is to conduct a thorough analysis of the existing literature and analyze insider threats from the perspectives of specialists in the field of cyber security.
The purpose of the research questions set above will pave the direction for which the systematic literature review will be carried out. Additionally, the research questions will also limit the scope of the study to prioritize literature to publications that were published recently as the world of cyber security is ever-revolving.
Apart from the number of studies that will be researched and studied on, case examples relating to insider threats will also be shown to examine how insider threats are able to impact the daily lives of individuals. This correlation will better put things into perspective to highlight the importance of insider threats.
At present, the literature review stage, our goal is to identify suitable literature to carry out an in-depth investigation to the topics and evaluate whether the research topics will be required to be further redefined or not.

2.1.	Selection of primary studies

Primary studies were conducted by passing keywords to the search facility or search engine. The Boolean AND and OR were used to conduct the search. The search terms were:
("All Metadata": “Insider Threats” OR "All Metadata": “Insider Threat detection” OR "All Metadata": “computer threats”) AND "All Metadata": “security”

(("All Metadata”: “Insider Threats” OR "All Metadata": “Insider Threat detection” OR "All Metadata": “Computer threats”) AND ("All Metadata": “cyber security” OR "All Metadata": “cybersecurity” OR "All Metadata": “cyber-security”)

The platforms searched were:
-	IEEE Xplore Digital Library
-	ScienceDirect
-	SpringerLink
-	ACM Digital Library
-	Google Scholar

Deep research was performed on titles, keywords or summaries, depending on the search platform. On October 9, 2022 We started looking for papers and we processed all studies published up to that date. The results of these searches are filtered using inclusion/exclusion criteria. At the end Snowball iteration forward and backward is performed to filter the papers. Only those papers are selected that meet the inclusion criteria.

2.2.	Inclusion and exclusion criteria

The research work chosen for this SLR should be only focused on insider threats, it’s detection and different techniques used to minimized or lower these attacks on the computer systems. Google Scholar, ACM Library, IEEE platform were searched to narrow down the papers. Those related to insider threats and detections were selected. The inclusion and exclusion criteria are shown in Table \ref{tab:Table 2}.

2.3.	Selection results

There was a total of 1323 studies identiﬁed by the initial keyword searches on the chosen platforms. After eliminating the duplicate research, this number was drastically reduced to 567. The research that met the inclusion/exclusion criteria were thoroughly examined, and 177 publications were left over for reading. Only papers written in English and published before 2020 were chosen. After reading all 177 papers in detail and using the inclusion/exclusion criteria once again, 50 papers were left for the SLR in the end.
\\
2.4.	Quality assessment

For the next step Five papers were selected randomly and were subjected to the following quality assessment process to check their effectiveness: 
\begin{table}
\caption{Inclusion and Exclusion criteria for the primary studies.}
\label{tab:Table 2}
\begin{tabular}{@{}ll@{}}
\toprule
Inclusion   Criteria & Exclusion   criteria                              \\ 
\midrule
\begin{tabular}[c]{@{}l@{}}The paper should focus on \\ the insider threats and its types\end{tabular}             & \begin{tabular}[c]{@{}l@{}}The papers should not be about \\ types of threats, it should focus \\ only on insider threats to the \\ cyber world.\end{tabular} \\
\begin{tabular}[c]{@{}l@{}}The paper should highlight the \\ insider detection techniques.\end{tabular}             & Papers published after 2020 
\\
\begin{tabular}[c]{@{}l@{}}The paper should focus on insider \\ threats challenges and its prevention\end{tabular} & \begin{tabular}[c]{@{}l@{}}The paper should focus on insider \\ threats challenges and its prevention.\end{tabular}                        \\ \bottomrule
\end{tabular}
\end{table}
\\

Stage 1: \textbf{Insider Threats:} The paper must be focused on the detection of insider threats and its prevention

Stage 2: \textbf{Context:} Adequate context for the research goal and results should be provided. This will allow for accurate study interpretation.

Stage 3:\textbf{ Insider threats detection:} The study must have sufficient data to accurately depict how the technology was used to solve a particular issue. which will eventually help in solving problem, RQ1 and RQ2.

Stage 4:\textbf{ Security context:} The paper must provide an explanation for security and privacy concern to answer the problem, RQ3.

Stage 5: \textbf{Performance:} The paper should provide enough evidence about performance of the system if the insider threat techniques are applied.

Stage 6: \textbf{Data acquisition:} Details about data collection, storing and filtering should be provided to assess how accurate it is.

All identified primary studies were then subjected to this checklist for quality assessment. Thirteen studies were found not to meet the one of the following checklists. So, they were removed from the list as shown in Table \ref{tab: Table 3}

\begin{table}[h]
\caption{}
\centering
\label{tab: Table 3}
\begin{tabular}{@{}ll@{}}
\toprule
Checklist for the Criteria Stages   & Excluded Studies \\ \midrule
Stage 1: Insider Threats            &  [S\ref{s47}]   , [S\ref{s46}] , [S\ref{s50}]            \\
Stage 2: Context                   &  [S\ref{s42}] , [S\ref{s45}]               \\
Stage 3: Insider threats detection & [S\ref{s39}] , [S\ref{s41}],  [S\ref{s44}]               \\
Stage 4: Security context           & [S\ref{s43}]                  \\ 
Stage 5: Performance                & [S\ref{s40}]  , [S\ref{s48}] ,[S\ref{s49}]              \\
Stage 6: Data acquisition           &     [S\ref{s38}]             \\ \bottomrule
\end{tabular}
\end{table}

2.5. Data extraction

Data were then extracted from all the articles that passed the quality assessment to assess the completeness of the data and verify the accuracy of the information contained in the articles. Before being deployed to all searches that passed the quality review stage, the data mining technique was first tested on years of preliminary investigations. Data from each of the study was extracted, classified into categories, and then entered into a excel sheet. The following categories were applied to the data:

\textbf{Context data}: Information on the study's purpose.

\textbf{Qualitative data}: The authors' findings and conclusions.

\textbf{Quantitative data}: Data obtained when we applied it to study and experiment.

2.6.  Data Analysis

Data corresponding to the qualitative and quantitative categories was assembled in order to answer all the research questions. The papers filtered after the completion of the data extraction process, were then passed through a meta-analysis process

\begin{figure}[h]
    \centering
    \includegraphics[scale=0.5]{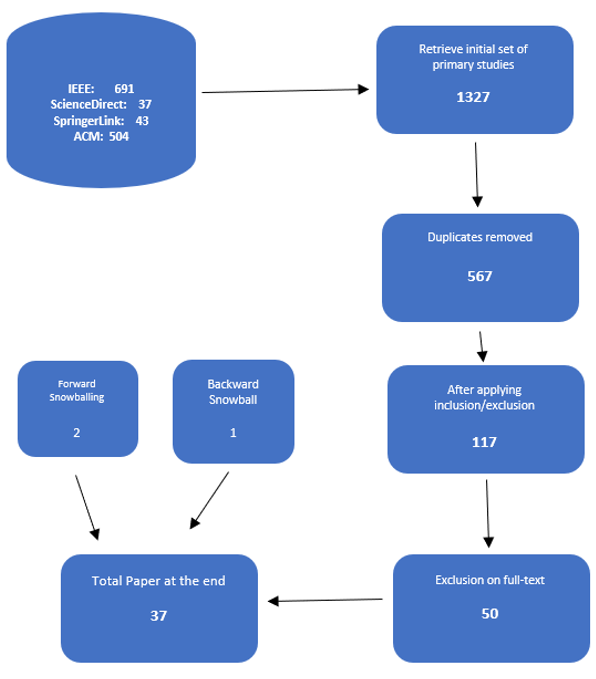}
    \caption{Attrition of papers through processing.}
    \label{fig:fig 1}
\end{figure}

Figure \ref{fig:fig 1} shows the number of papers selected as the initial result and it also represents the process before selecting the number of final papers.

\begin{figure}[h]
    \centering
    \includegraphics[scale=0.5]{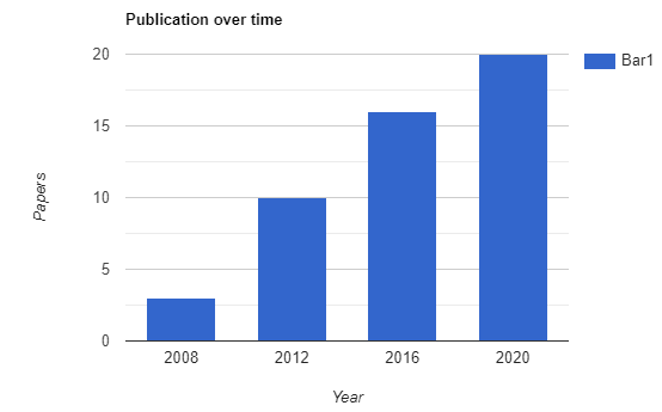}
    \caption{Number of primary studies published over time}
    \label{fig:fig 2}
\end{figure}

1.	Publications over time: As it is clear from the graph in Figure \ref{fig:fig 2} that in recent years, there has been significant increase in the Publication of papers based on cyber security and insider threats. This is because recent many attacks related to this insider techniques have been occurred globally. This shows that with modernization of internet, increase in threats and cyber security has also boomed in the global market.

2.	Significant Keyword Count: An analysis of keywords was performed across all 37 studies in order to summaries the common themes among the selected primary studies. Table \ref{tab:Table 4} shows how many times certain words appeared in all of the primary studies. As shown "Insider Attack" is the most frequently occurring keyword in our dataset, following "network" and " Threats" \cite{a6,a7}.

\begin{table}[h]
\centering
\caption{}
\label{tab:Table 4}
\begin{tabular}{@{}ll@{}}
\toprule
Keywords          & Count \\ \midrule
Insider Attack    & 1521  \\
Network           & 1201  \\
Threats           & 855   \\
Machine Learning  & 724   \\
National security & 356   \\
Humans            & 250   \\
Cyber             & 235   \\
Internet          & 183   \\ \bottomrule
\end{tabular}
\end{table}

\section{Findings}
\label{Findings}
\begin{figure}
    \centering
    \includegraphics[scale=0.3]{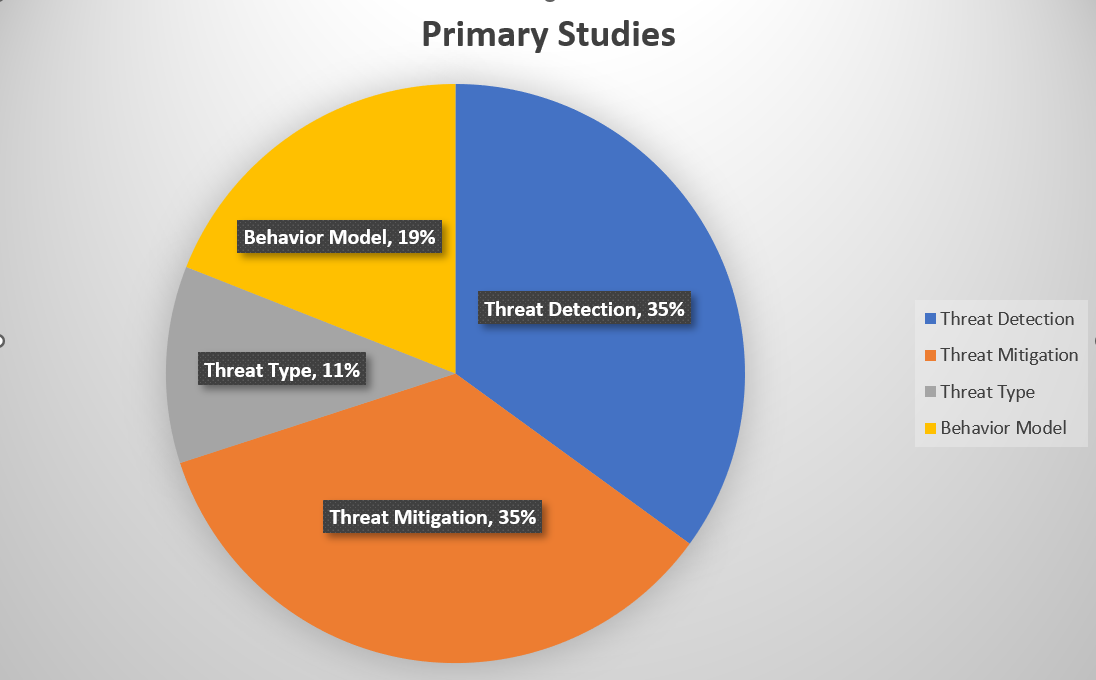}
    \caption{Chart of themes of primary studies}
    \label{fig:fig 3}
\end{figure}

Each primary research study was read thoroughly and relevant qualitative and quantitative information was taken out and summarised in Table \ref{tab:Table 5}. All the papers has basic information about threat types and its detection. The main theme of paper is also summarized at the end.
Each paper was subdivided into broader categories. It is clear from Fig \ref{fig:fig 3}, that majority of studies were focused on threat detection and threat mitigation in insider threats. 
Fig \ref{fig:fig 3}, shows the graphical representation of subdivided categories and their applications in insider threats.
The themes found in primary studies shows that threat detection and threat mitigation are the most popular security application in insider threats, with both comprising of 35 percent each respectively. These two applications comprise the majority of the studies section. Threat detection discus about, the detection techniques adopted by companies to track insider threats. Whereas, threat mitigation refers to elimination of the causes of threats in a security system. The third most popular among the primary studies was, modeling human behaviors to detect the insider threat attack (behavior model). It comprises of 19 Percent of the total studies. At the last comes the types of threat, which makes up 11 Percent of the total studies.

\section{Discussion}
\label{Discussion}
Initial keyword searches reveal that there are a large number of papers on Insider Attack. This is due to reason that with advancement in internet more and more social media interaction is causing these attacks. As people are easily convinced on SM and they reveal their secrets to others people and company.

\begin{table}[]
\caption{Main findings and themes of the primary studies.}
\label{tab:Table 5}
\scalebox{0.8}{
\begin{tabular}{@{}|l|l|l|@{}}
\toprule
\    
{    \begin{tabular}[c]{@{}l@{}}Primary \\ Study\end{tabular}} & {    \begin{tabular}[c]{@{}l@{}}Key Qualitative and Quantitative \\ Data Reported\end{tabular}} & {    \begin{tabular}[c]{@{}l@{}}Types of Security \\ Applications\end{tabular}} \\ \midrule
\    
{  [S\ref{s1}}] & {    \begin{tabular}[c]{@{}l@{}}The author focuses on the insider threats   \\ and defense measures against them. \\ They used top 100 papers and identified \\ five different defense solution approaches.\end{tabular}} & {    Threat Detection}                                                          \\ \midrule
{    [S\ref{s2}}] & {    \begin{tabular}[c]{@{}l@{}}This paper mentions insider threats \\ detection using behavior modeling \\ and algorithms. The author uses \\ log-based data like, browser history,\\ email contacts to detect the insider threats.\end{tabular}}        & {    Behavior Model}                                                            \\ \midrule
\    
{    [S\ref{s3}}] & {    \begin{tabular}[c]{@{}l@{}}Use of machine learning algorithms \\ and mechanism   was suggested in \\ this study by the authors. The authors \\ divided the insider   behaviors into four \\ different classes and suggested to research \\ more on each  induvial class with help \\ of machine learning algorithms.\end{tabular}} & {    Behavior Model}                                                            \\ \midrule
\    
{    [S\ref{s4}}] & {    \begin{tabular}[c]{@{}l@{}}The paper discusses on how Insider Threats\\ has impacted our critical infrastructure like \\ healthcare and the effectiveness of \\ current defensive measures.\end{tabular}}  & {    Threat mitigation}                                                         \\ \midrule
\    
{    [S\ref{s5}}] & {    \begin{tabular}[c]{@{}l@{}}The author describes the insider attacks \\ and data exfiltration. This study highlights, \\ survey on the data leakage   patterns and \\ countermeasures regarding it moreover, \\ author also mentions the   three different \\ data states and how often they are attacked.\end{tabular}} & {    Threat Type}                                                               \\ \midrule
\    
{    [S\ref{s6}}] & {    \begin{tabular}[c]{@{}l@{}}Cyber-attacks have been a main cause of \\ concerns for companies over years. \\ But majority of them are done by the \\ insiders. Author has defined different \\ types of insiders and attacks technique \\ used by them to carry these malicious activities.\end{tabular}} & {    Threat Type}                                                               \\ \midrule
\    
{    [S\ref{s7}}] & {    \begin{tabular}[c]{@{}l@{}}SCADA systems are prominent in critical \\ infrastructures and preventing them from \\ cyber-attack is important issue. It reviews \\ the current strategies like honey pots used \\ in SCADA and their efficiency.\end{tabular}}     & {    Threat mitigation}    
\\ \midrule

{    [S\ref{s8}}]                                                    & { \begin{tabular}[c]{@{}l@{}}With advancement of technology, insider \\ attacks are also becoming advanced. \\ This study investigates different types \\ of risk perception models and its characteristics. \\ A fair solution has been proposed by the \\ author at the end of this study.\end{tabular}}  
   & {    Behavior Model}  \\ \midrule                                                                                                                         

{    [S\ref{s9}}]  & {    \begin{tabular}[c]{@{}l@{}}This article reviews the involvement \\ of humans in insider   attack, \\ and how it really affects the other \\ employees in the group. The author \\ defines how it can trigger the shift in \\ trustworthiness within fellow group \\ member or within the company.\end{tabular}}         & {    Behavior Model}                                                            \\ \midrule
\    
{    [S\ref{s10}}] & {    \begin{tabular}[c]{@{}l@{}}The author explains a famous method \\ known   as DLPS (data leakage \\ prevention system) and how it is \\ being utilized by IT   companies to \\ protect data breach and leakage. \\ DLPS use different methods to prevent \\ the data leak and author has described \\ its weakness as well as methods to improve them.\end{tabular}}                                   & {    Threat Mitigation}                                                         \\ \midrule
\    
{    [S\ref{s11}}]                                                   & {    \begin{tabular}[c]{@{}l@{}}Here, author describes the newer concept \\ known as Information Security (InfoSec) \\ research, i.e., approaches to prevent \\ intrusion in company system. \\ The study concludes the approaches \\ to detect and mitigate threats using InfoSec.\end{tabular}}                  & {    Behavior Model}                                                            \\ \midrule
\    
{    [S\ref{s12}}]                                                   & {    \begin{tabular}[c]{@{}l@{}}In this paper, author introduced a new and \\ more efficient way of dealing with insiders. \\ Author divided the threats into two main parts; \\ intentional and accidental. Author suggested \\ to use a new evaluation method based on \\ the profile and position of the users in \\ the system or organization to predict the threats.\end{tabular}}                   & {    Threat Mitigation}                                                         \\ \midrule
\    
{    [S\ref{s13}}]                                                    & {    \begin{tabular}[c]{@{}l@{}}In this author reviews insider behavior \\ and their role in harming data, whether \\ it’s on organization level or cloud level. \\ Author discusses its impact and benefits \\ insider gets from it.\end{tabular}}                    & {    Threat Mitigation}                                                         

                   \\ \bottomrule
\end{tabular}}
\end{table}

\begin{table}
\ContinuedFloat

\caption{Continued}
\label{tab:table 5}
\scalebox{0.8}{
\begin{tabular}{@{}|l|l|l|@{}}
\toprule
\    
{    \begin{tabular}[c]{@{}l@{}}Primary \\ Study\end{tabular}} & {    \begin{tabular}[c]{@{}l@{}}Key Qualitative and Quantitative \\ Data Reported\end{tabular}} & {    \begin{tabular}[c]{@{}l@{}}Types of Security \\ Applications\end{tabular}} \\ \midrule
\    
{    [S\ref{s14}}]                                                    & {    \begin{tabular}[c]{@{}l@{}}Author suggested that single authentication \\ is not efficient method for security. \\ So, the study proposed a new re-authentication \\ method using ensemble technique. \\ Where, author uses keystroke methods \\ to record a pattern to verify the user is \\ authentic or not.\end{tabular}}                                                                                                                                                                                                                                                                                                                                                                                             & {    Threat Mitigation}                                                         \\ \midrule
\    
{    [S\ref{s15}}]                                                    & {    \begin{tabular}[c]{@{}l@{}}In most organization, most of attacks are \\ done by insiders and which are hard to detect. \\ Author defined a system known as Internal \\ Intrusion Detection and Protection \\ System (IIDPS), where system-call patterns \\ were used to record log data and a profile \\ is created according   to that.\end{tabular}}                                                                                                                                                                                                                                                                                                                                                                    & {    Behavior Model}                                                            \\ \midrule
\    
{    [S\ref{s16}}]                                                   & {    \begin{tabular}[c]{@{}l@{}}In This article, author presents an innovative \\ detection method that is robust against \\ mimicry-based evasion strategies. The author \\ divided processing of information into two \\ different stages. First   one is analysis stage, \\ where user actions are recorded and then \\ in second stage they are verified.\end{tabular}}                                                                                                                                                                                                                                                                                                                                                    & {    Threat Detection}                                                          \\ \midrule
\    
{    [S\ref{s17}}]                                                   & {    \begin{tabular}[c]{@{}l@{}}Author defined that using single model \\ methods to detect insider attacks is not efficient. \\ Study shows that using ensemble technique \\ to detect the threats is better than single model. \\ The author used both of the supervised and \\ unsupervised learning techniques to detect the insider. \\ It came out that supervised learning is \\ better than unsupervised learning.\end{tabular}}                                                                                                                                                                                                                                                                                       & {    Threat mitigation}                                                         \\ \midrule
\    
{    [S\ref{s18}}]                                                    & {    \begin{tabular}[c]{@{}l@{}}This survey paper mentions the SCADA\\  system Vulnerabilities and technique \\ named alarm-and-trust-based access \\ management system (ATAMS). \\ ATAM detect user performance periodically \\ and keeps an eye if an anomaly is detected.\end{tabular}}                                                                                                                                                                                                                                                                                                                                                                                                                                     & {    Threat Detection}                                                          \\ \midrule
\    
{    [S\ref{s19}}]                                                    & {    \begin{tabular}[c]{@{}l@{}}This paper continues the discussion of the \\ risks posed by Hardware like USB devices. \\ Hardware trojan concept is used by \\ author to filtrate data from USB hardware.\end{tabular}}                                                                                                                                                                                                                                                                                                                                                                                                                                                                                                      & {    Threat Detection}                                                          \\ \midrule
\    
{    [S\ref{s20}}]                                                   & {    \begin{tabular}[c]{@{}l@{}}This paper mentions the forensic \\ investigation technique to prevent \\ threat attacks. This technique has two \\ phases: first phase is monitoring \\ incoming and outgoing packets and \\ data they contain. Where as in phase \\ two: critical log data is verified for \\ any sensitive information.\end{tabular}}                                                                                                                                                                                                                                                                                                                                                                       & {    Threat Mitigation}                                                         \\ \midrule
\    
{    [S\ref{s21}}]                                                    & {    \begin{tabular}[c]{@{}l@{}}Because insiders of a company have \\ access to sensitive information, \\ insider attacks can be devastating. \\ This article proposes a novel insider \\ attack authentication protocol based \\ on the dependable cryptographic \\ algorithm known as ECC. It \\ demonstrates that the protocol has \\ a high level of security while being \\ relatively better than other \\ existing protocols in terms of \\ computational cost and communication \\ overhead.\end{tabular}}                                                                                                                                                                                                             & {    Threat Detection}                                                          \\ \midrule
\    
{    [S\ref{s23}}]                                                    & {    \begin{tabular}[c]{@{}l@{}}Collaborative intrusion detection \\ networks (CIDNs), allows an IDS to \\ gather data and pick up expertise from others. \\ it has been created to improve the \\ detection accuracy in order to fight \\ against complicated attacks. However, \\ these networks are prone to security issues. \\ therefore, author created an intrusion \\ sensitivity-based trust management \\ model that enables each IDS to assess \\ the trustworthiness of others by checking \\ their detection sensitivity.  Author used \\ trust model in a real wireless sensor  \\ network and compared the effectiveness \\ of three different supervised classifiers \\ under attack conditions.\end{tabular}} & {    Threat Detection}                                                          \\ \midrule
\    
{    [S\ref{s24}}]                                                    & {    \begin{tabular}[c]{@{}l@{}}The Internet of Things (IoT) is a new \\ approach in computer networks that \\ uses various technologies to connect \\ different objects all around us to each \\ other and Internet. The author introduced \\ a new model for IOT that provides \\ with intrusion detection agents based on \\ both supervised and unsupervised \\ learning and has ability to detect \\ both internal and external attacks \\ simultaneously.\end{tabular}}                                                                                                                                                                                                                                                  & {    Threat Detection}                                       
                   \\ \bottomrule
\end{tabular}}
\end{table}
\begin{table}
\ContinuedFloat
\caption{Continued}
\scalebox{0.8}{
\begin{tabular}{@{}|l|l|l|@{}}
\toprule\
\ 
{    \begin{tabular}[c]{@{}l@{}}Primary \\ Study\end{tabular}} & {    \begin{tabular}[c]{@{}l@{}}Key Qualitative and Quantitative \\ Data Reported\end{tabular}} & {    \begin{tabular}[c]{@{}l@{}}Types of Security \\ Applications\end{tabular}} \\ \midrule
\
{    [S\ref{s24}}]                                                   & {    \begin{tabular}[c]{@{}l@{}}During data transfer, there are two types of data loss. \\One is link error and another is malicious packet drop.\\ Data lost by malicious activity is always overlooked\\ by researchers. In this paper, author has deeply mentioned\\ and studied the packet drop by malicious activities. Author \\ developed a homomorphic linear authenticator (HLA) based\\ public auditing architecture that verifies the truthfulness \\of packet loss.\end{tabular}}                                                                                                                                                                                                                & {    Behavior Model}                                                            \\ \midrule
\    

{    [S\ref{s25}}]                                                   & {    \begin{tabular}[c]{@{}l@{}}In order to support a conventional type of trust \\evaluation that includes evidence summing, Author suggest \\ two workable techniques to protect the privacy of trust \\evidence providers based on additive homomorphic encryption.\\ The first scheme achieves superior computational efficiency. \\ While the second approach offers stronger security.\end{tabular}}                                                                                                                                                                                                                                                                                                      & {    Threat Mitigation}                                                         \\ \midrule
\    
{    [S\ref{s26}}]                                                    & {    \begin{tabular}[c]{@{}l@{}}This paper offers a defensive deception strategy that is \\OS-resident and can be used to neutralize malware \\infected machine.\end{tabular}}
& {    Threat Detection}                                                          \\ \midrule
\    
{    [S\ref{s27}}]                                                   & {    \begin{tabular}[c]{@{}l@{}}One of the main obstacles to the successful implementation \\of fog computing and the Internet of Things (IoT) is device\\ security. Author described, a three technologies \\ strategy that includes a Markov model, an intrusion detection \\system (IDS), and a virtual honeypot device (VHD) to \\recognize malicious devices with the \\help of fog computing environment\end{tabular}}                                                                                                                                                                                                                                                                                                 & {    Threat Detection}                                                          \\ \midrule
\    
{    [S\ref{s28}}]                                                    & {    \begin{tabular}[c]{@{}l@{}}The Article explains the fundamentals of the fog computing\\ technology system, outlines three security issues—creation \\and use of dummy documents, unidirectional \\transparent detection—and suggests solutions.\end{tabular}}                                                                                                                                                                                                                                                                                                                                                                                                                                                    & {    Threat Detection}                                                          \\ \midrule
\    
{    [S\ref{s29}}]                                                    & {    \begin{tabular}[c]{@{}l@{}}In this study, Author suggested a smart reaction mechanism \\that constantly manages the operational risks to human\\ operators. Additionally, the architecture  of this mechanism \\enables the system to assume control when areas are \\completely unprotected and/or isolated in addition to \\allowing the system to analyses the status of a particular \\scenario on its own.\end{tabular}}                                                                                                                                                                                                                                                                                & {    Threat Mitigation}                                                         \\ \midrule
\    
{    [S\ref{s30}}]                                                    & {    \begin{tabular}[c]{@{}l@{}}The author discusses an architecture for overlay networking \\that is based on gossip protocols, which enable users to \\communicate with one another other in a safer way.\\ A characteristic of the design is that it limits the quality and\\ quantity of information that insiders may access while\\ optimizing the routing of requests to only include individuals \\who can respond to them.\end{tabular}}                                                                                                                                                                                                                                                                                        & {    Threat Mitigation}                                                         \\ \midrule
\    
{    [S\ref{s31}}]                                                    & {    \begin{tabular}[c]{@{}l@{}}Author developed a management protocol that   prevents \\both outsider and insider security attacks for group\\ communication systems (GCSs) in mobile ad hoc \\ networks (MANETs).\end{tabular}}                                                                                                                                                                                                                                                                                                                                                                                                                                                                                         & {    Threat Mitigation}                                                         \\ \midrule
\    
{    [S\ref{s32}}]                                                   & {    \begin{tabular}[c]{@{}l@{}}In this study, Author created a software package that could \\be set up as a gateway between users and the legacy servers. \\ This software uses TCP tunnelling and a hardware device \\that was scaled down to detect insider attacks.\end{tabular}}                                                                                                                                                                                                                                                                                                                                                                                                                                  & {    Threat Type} 
                                       \\ \midrule
\  
{    [S\ref{s33}}]                                                    & {    \begin{tabular}[c]{@{}l@{}}The author mentions, framework based on process models\\ for real-time log data analysis as it is impossible for \\human auditors   to track it.\end{tabular}}                                                                                                                                                                                                                                                                                                                                                                                                                                                                                                                            & {    Threat Detection}                                                          \\ \midrule
\ {    [S\ref{s34}}]                                                    & {    \begin{tabular}[c]{@{}l@{}}The primary objectives of this work are to identify and \\forecast assaults  and harmful behaviors through the \\analysis, classification, and labelling of activities that have\\ been recorded in log files. In this study, each user's\\ behavior is predicted using MapReduce programming. \\ Unknown events are clustered using the K-Means approach, \\and unlabeled classes are defined using K-NN supervised \\ learning on the NSLKDD database.\end{tabular}}                                                                                                                                                                                                            & {    Threat Type}   
                                   \\ \midrule
\    
{    [S\ref{s35}}]                                                    & {    \begin{tabular}[c]{@{}l@{}}Author described an approach that can be applied to the \\challenge of detecting insider threats, specifically the \\exploitation of an organization's resources by system users.\end{tabular}}                                                                                                                                                                                                                                                                                                                                                                                                                                                                                        & {    Threat Detection}                                                          \\ \midrule
\    
{    [S\ref{s36}}]                                                    & {    \begin{tabular}[c]{@{}l@{}}Author described application of machine learning in \\Threats. Author used CERT Insider Threat Dataset and\\ applied ML algorithm on them to calculate the results.\end{tabular}}                                                                                                                                                                                                                                                                                                                                                                                                                                                                                                      & {    Threat Mitigation}                                                         \\ \midrule
\    
{    [S\ref{s37}}]                                                   & {    \begin{tabular}[c]{@{}l@{}}In this study, Author introduced the community anomaly \\detection system (CADS), a framework for unsupervised\\ learning that uses access logs from collaborative environments \\to identify insider threats.\end{tabular}}                                                                                                                                                                                                                                                                                                                                                                                                                                                         & {    Threat Detection}                           

                   \\ \bottomrule
\end{tabular}
}
\label{}
\end{table}
\
Wide range of papers discuss the threat detection and its removal techniques. insiders are familiar with an organization's systems; their malicious behavior is impossible to detect. Therefore [S\ref{s2}], offer a unique way of detecting and minimizing it. The author, proposes an insider threat detection method based on user behavior modeling. The user's daily activity, the delivery of email content topics, and the user's weekly email communication history comprises of common log activities used for detection. 
As mentioned earlier, in order to protect insider attacks, user behavior as well as hackers’ behavior should also be kept in mind [S\ref{s11}]. Every corporation, regardless of size, requires safe IT. Log management is always the critical step while managing firewalls. Article [S\ref{s33}] and [S\ref{s34}] describes how real-time log data analysis is important and how it can be tracked. The log files that are kept on servers, firewalls, and other IT hardware serve as a record of significant transactions and events. This data may offer crucial hints about malicious behavior affecting your network both inside and externally. The analysis of log data can also help in locating and resolving hardware failures as well as setup issues with equipment.
Many researchers are shifting their approaches toward the use of Machine Learning for threat Management. In [S\ref{s35}] researchers used CERT Insider Threat Dataset and applied ML algorithm on them. The authors of this study develop a framework called RADISH in and proved that it is capable of detecting insiders using both -Nearest Neighbor (-NN) and -Dimensional Tree (- tree) against the CERT r2 dataset. According to their findings, -NN is often quicker and more accurate than -d tree.
Due to this, primary studies have shown rising trend of use of Machine learning in Insider attack detection. Moreover, most of the study pointed that behavior modeling should also be considered. This includes insiders’ behavior also, as most the organization neglect their behavior and end up losing data and resources.\\

\textbf{RQ 1: What is the magnitude and impact of insider threats to businesses globally?}

It is common that, when a person thinks of insider threats, disgruntled and unethical employee comes in our mind. However, there is also a chance that an uninformed employee or contractor could expose the company network to danger by giving privileged credentials to an outside threat actor who then uses them to impersonate an insider. The average global cost of insider threats rose to 11.45 US million dollars in the last two years, according to the 2020 Cost of Insider Threats Global Report, and the number of occurrences rose by 47 percent during that time \cite{o5}. Therefore, the economic impact of these attacks is severe and countermeasures should be seriously considered. Insider threat has been an issue for national security and global business also. As, we know that technology advancement has boosted the E-Payment globally, it becomes more critical to understand this as the software system advancements brings cyber trouble also. For Example: In February 2016, Bank of Bangladesh, hackers tried to rob 1 billion dollars and they almost succeeded in doing this. They almost attempted to do the biggest Cyber-attack of the history  \cite{o6}.\\

\textbf{RQ 2: The common techniques used to carry out insider threats and override controls that are set to detect insider threats in the first place.}

Despite all the work and money put into developing cybersecurity threat detection systems, detecting a malicious activity can take weeks or even more than that.  Like, in Bangladesh robbery case, the hackers are still unidentified \cite{c2,a9}. However, taking a deep analysis of how an attack is done can help in detecting a cyber-attack. The reason why it’s hard to detect a cyber attack is, because most of these are caused by trusted user or ex- employee, who has access to data and resources. According to a data survey from a website, the cost to handle an insider threat is around 8 million dollars on an average \cite{o7}. There are numerous ways to carry out insider threats. But the main three of them are:
\\
i.	Data Theft
\\
ii.	Privilege Exploitation
\\
iii.	Sabotage
\\
\textbf{I.	Data Theft:} it mainly the most popular insider threat widely. There can be numerous reasons for this attack. Ex-employee trying to defame company, Anti company trying to push your company down. The following indicators can be used to detect these types of attacks at first place:
\\
i.	Accessing data when employee is not working.
\\
ii.	Using company’s interface on odd device or your personal device.
\\
iii.	Downloading data into your own device.
\\
\textbf{II.	Privilege Exploitation:} When a employee has access to sensitive information about an organization or company. They can use that information to bring it done or leak it outside. However, it’s hard to detect these but there are following ways with which it can be done:
\\
i.	Altering organizations privileges
\\
ii.	Adding new users
\\
iii.	Changing security authentications without any permission.
\\
\textbf{III.	Sabotage:} This can be defined as a revenge from an employee for being treated unfair or blackmailing the company. This can be detected by following ways:
\\
i.	Constantly keeping an eye on Managers and colleagues’ behavior towards another employee.
\\
ii.	Recent argument between employee and company executives
\\
iii.	Requesting access for the privileges that user doesn’t need.\\
\\
\textbf{RQ 3: What methods and techniques are available to overcome and insider threats in the world of cyber security?}
\\
Although, it is hard to eradicate the insider threats issues, but we can use number of ways to minimize it. The fact that we are frequently betrayed by the people we most trust. But still company has to keep faith in their people to run their business smoothly. However, they can use these methods to minimize these and keep track of the attacks.
\\
i.	Company should follow a strict security policy rule, and should form a department that keeps the track of the malicious activities inside and outside the system.
\\
ii.	Infrastructure should be locked with proper security and employee should be given secured lockers to lock their physical and sensitive information and documents.
\\
iii.	The main cause of these activities are mostly the new hires, company should spend money and screening new hires and they shouldn’t be given the full authority in the beginning of their job.
\\
iv.	We shouldn’t blame humans always for these attacks, our computer system should be secured with auto shutdown system when there is malicious activity or data breach.
\\
v.	Direct employee monitoring is a further helpful tool. You can never be too safe when it comes to your company's confidential information. This can be done using security cameras or implementing keystroke logging.
\\
We can increase the security of the confidential data at our company by putting these insider threat detection techniques into practice. If company is not putting these techniques into action, they may suffer loss of thousands of dollars due to these continuous cyber-attacks. Specially, those in which the insiders are involved and data leak is from inside rather than outside activity.

\section{Future Research Directions of Insider Threats}
\label{Future Research Directions for Insider Threats}
Based on our research, we presented the following research directions for Insider threats in cyber security.

\textbf{i.	Threat Type in Cyber Security:} As type of threat, we are dealing with plays a key role in cyber security \cite{c1}. If the threat  we are dealing is caused by the insider then it is hard to detect and caused high damage to organization also. It is clear that almost every article describes the threat type, but not all of them were able to explain the detection method. As it can be observed that if type of threat is insider, then the loss to company is also high. For example: in case of bank of Bangladesh.

\textbf{ii.	Behavior Model in Cyber Security:} Modeling the behavior of the user and employee is the another key factor that can help to minimize threats in future. This includes keeping track of user email activity, log books and their behavior in the organization.

\textbf{iii.	Threat Detection:} Threat detection helps a company to get a brief info about past attacks and vulnerabilities in the system. We have discussed many techniques used to detect these attacks. Threat detection refers to an IT organization's ability to quickly and accurately detect threats to a network, its applications, or other assets. Understanding the threats that exist in the cyber environment is the first step in developing efficient threat detection and response processes. Organization should have a committee which deals with these attacks and who are ready to take action. So that in future these attacks can be minimized.

\textbf{iv.	Threat Mitigation:} Threat detection are of no use if organizations are not taking measures to eradicate it. As, discussed in RQ3, company should screen new hires, keep log books of employee data, should follow multi factor authentication. These potential steps can easily help to decrease future cyber attacks on company

\section{Conclusion and Future Work}
\label{Conclusion and Future Work}
This research main focus was on insider threats and its detection. Although, we covered a brief part about cyber security also. The paper highlights the main types and the methods that we can use to minimized those attacks.
This research mentions the new discovered methods that has never been used by companies for attack mitigation. Early detection of threats has always been helpful for companies, but most the companies neglect it and loose data and money. Although, it is impossible to completely make a company or system attack-free. Many developments have been made in this field already and more is under progress. The research highlights the potential for future work and the main three agenda are discussed below:

\textbf{Potential Research Agenda 1:} Researchers has offers numerous ways of dealing with these insider attacks. However, Use of behavior model techniques to keep an eye on users can be the one of the most efficient techniques for threat detection. Although, researchers still lack in this area as not enough data is available regarding this. Future work could include more deep assessment of human behavior patterns and attacks strategy they use.

\textbf{Potential Research Agenda 2:} Several of the study highlighted that insiders’ attacks are mainly by the ex- employee or an unhappy user. Future work could include a review of how and why these insiders are influenced to do such things.

\textbf{Potential Research Agenda 3:} With the rising of internet and digital world, data transfer is also doubling every year. Out of which, more than half is vulnerable to loss or attacks. Hence, a potential research agenda could be protecting the rising perimeter of internet. For example, to be fully prepared for the worst, organization should invest in comprehensive risk management and compliance solutions. Companies can stay one step ahead of attackers and avoid the worst from ever happening.

\section{Declarations of interest}
\label{Declarations of interest}

None.
\section{Acknowledgement}
\label{Acknowledgement}

I would also like to extend my deepest gratitude to Professor Ali Dehghantanha for their assistance in completing the project.

\

\bibliographystyle{ieeetr}
\bibliography{main} 

\renewcommand\refname{Primary Studies}

\end{document}